\documentclass[prd,twocolumn,showpacs,nofootinbib,preprintnumbers,amsmath,amssymb]{revtex4}
\usepackage[dvipdfm]{graphicx}
\usepackage{amssymb}
\usepackage{amsmath}
\usepackage{parskip}
\usepackage{epsfig}
\usepackage{dcolumn}
\usepackage{rotating}
\newcommand{\bea}{\begin{eqnarray}}
\newcommand{\eea}{\end{eqnarray}}
\newcommand{\be}{\begin{equation}}
\newcommand{\ee}{\end{equation}}
\newcommand{\beqy}{\begin{eqnarray}}
\newcommand{\eeqy}{\end{eqnarray}}
\newcommand{\p}{\partial}

\newcommand{\mx}{\mbox}
\newcommand{\mt}{\mathtt}

\newcommand{\al}{\alpha}
\newcommand{\bb}{\beta}

\newcommand{\Ga}{\Gamma}

\newcommand{\de}{\delta}
\newcommand{\De}{\Delta}

\newcommand{\om}{\omega}
\newcommand{\Om}{\Omega}

\newcommand{\ti}{\tilde}

\newcommand{\taut}{\ti{\tau}}

\newcommand{\cH}{{\cal H}}
\newcommand{\cO}{{\cal O}}

\newcommand{\ra}{\rightarrow}
\newcommand{\Ra}{\Rightarrow}
\newcommand{\im}{\Longleftrightarrow}

\newcommand{\LF}{\left(}
\newcommand{\RF}{\right)}
\newcommand{\LT}{\left[}
\newcommand{\RT}{\right]}
\newcommand{\mtx}{\mt{max}}
\newcommand{\mtc}{\mt{eq}}
\newcommand{\mte}{\widetilde{\mt{eq}}}
\newcommand{\mth}{\widetilde{\mt{hub}}}
\newcommand{\ie}{{\it i.e.\ }}

\begin{document}
\preprint{IGC-08/12-1}
\title{Cyclic Inflation}
\vspace{3cm}
\author{Tirthabir Biswas$^a$}
\email{tbiswas@gravity.psu.edu}
\vskip 3cm
%\begin{center}
\author{Stephon Alexander$^{ab}$}
\email{sha3@psu.edu}
\vskip 3cm
%\begin{center}
\affiliation{
{\it
Department of Physics\\
Institute for Gravitation and the Cosmos\\
The Pennsylvania State University\\
104 Davey Lab, University Park, PA,16802, U.S.A \\}
\vskip 1mm
}
\affiliation{ {\it $^b$ Department of Physics and Astronomy\\ Haverford College\\ Haverford, PA,19041, U.S.A \\} }

%\end{center}

%\maketitle

\date{\today}

\begin{abstract}
We present an inflationary model that is geodesically complete and does not suffer from the transplanckian problem.  In most inflationary models, massless (conformal) scalar field fluctuations in a  deSitter background gives rise to a scale-invariant spectrum. In this work, we realize scale invariant perturbations from thermal fluctuations in (conformal) radiation during a  radiation dominated contraction era prior to inflation. As the modes exit the Hubble radius during the contraction phase, scale invariant fluctuations are indeed generated.  After many cycles, we enter into a power-law inflationary phase, that stretches the modes produced in the previous contraction phase to scales that we observe today.
\end{abstract}

%\pacs{98.80.Cq,98.80.Es,74.20.Fg}% Inflation, Cosmological Constant, and BCS Theory

\maketitle

%\newpage

%\setcounter{page}{1}
%%%%%%%%%%%%%%%%%%%%
\section{Introduction}
While the inflationary paradigm is successful in resolving the classic puzzles of the Standard Big Bang (SBB) cosmology and providing a mechanism for generating near scale-invariant fluctuations, it suffers from specific theoretical/technical issues. {\bf (i)} Perhaps the most daunting is the fact that inflationary space-times are geodesically incomplete and, hence, do not address the issue of the BB singularity~\cite{vilenkin}. ({\bf ii}) In many inflationary scenarios  one encounters the transplanckian problem~\cite{transplanck} where the scales one observes today becomes smaller than the Planck length at the beginning of inflation; it is not clear that one can trust the assumption of perturbation theory in the transplanckian regime  (see~\cite{transplanck,trans} for a few different paradigms where attempts have been made to quantify the corrections). {\bf (iii)} It is typically challenging to build a phenomenologically viable inflationary model where the potential remains flat for a long enough period. {\bf (iv)} Finally, it has been realized for some time now that inflation initiates in a region which is  extremely low in entropy.  For inflation to work, although  one only requires a microscopic ``smooth'' patch where the scalar field is homogeneous and isotropic, a priori, in the context of a continuum field theory this is not guaranteed~\cite{low-entropy};  consequently, it has been argued that the ``initial conditions'' necessary to begin inflation may not be ``sufficiently likely'', although there are counter-arguements~\cite{linde-ic} and proposals~\cite{way-around} for getting around this difficulty.  In this paper we explore an extension of inflation that addresses most of the aforementioned problems.

Cyclic models offer an interesting alternative to the inflationary paradigm; they evade the conceptual and technical problems with a ``beginning of time'' when the universe starts from a singularity by positing time to be ``eternal'' in either direction (past and future)\footnote{It is rather surprising to note that in the context of four dimensional isotropic and homogeneous cosmology, apart from having $\spadesuit$ an infinite number of cycles, there are only three other viable non-singular alternatives in the past: $\heartsuit$ the bouncing universe where there is a single bounce and our current phase of expansion is preceded by a phase of contraction as for instance advocated in pre-big bang scenarios~\cite{pre-bigbang} and non-local gravity models~\cite{bms}. For a related interesting approach where at the bounce point the arrow of time reverses,  see~\cite{moffat}. $\diamondsuit$ the emergent universe scenario where the universe asymptotes to a constant space-time in the infinite past~\cite{emerge}. $\clubsuit$ The universe has  a finite number of cycles where the first one begins either as $\heartsuit$ or $\diamondsuit$.}. This alternative view has motivated physicists to revisit cyclic universes time and again~\cite{tolman,narlikar,ekcyclic,barrow,dabrowski,phantom,kanekar,saridakis} (for a comprehensive review see~\cite{review}). Further, most cyclic models automatically address some of the classic problems of Big Bang (BB) cosmology which were the motivations for inventing inflation in the first place.  Thus although cyclic and inflationary models are not mutually exclusive, it is natural to try to attempt to replace inflation altogether with ``cyclicity''.  In this paper, however, we take a slightly different approach, by exploring whether by embedding inflation in a cyclic universe setting, some of it's problems viz. {\bf (i-iv)} can be alleviated.

Our main idea is to merge inflation with cyclic cosmology where the universe undergoes an infinite number of cycles before bouncing into a final power-law inflationary phase.  However, unlike conventional inflation where one exploits the fact that massless scalar field fluctuations in a conformally invariant deSitter background give rise to a scale-invariant spectrum, in this model a different ``conformal phase'' is employed to render scale-invariant fluctuations.  It is quite remarkable that these fluctuations generically produce a scale-invariant spectrum, unlike inflationary or ekpyrotic models which rely on special kind of  scalar field potentials. First alluded to by Peebles~\cite{peebles}, this was elaborated on later in the context of a bouncing universe in~\cite{pogosian}. The problem with this scenario  is that in a symmetric bounce the fluctuations corresponding to cosmic structures that we see today is generated at very low temperatures (during contraction) and hence  are extremely suppressed. This is where the power-law inflationary phase comes into play. As is well known, inflation can stretch  fluctuations generated at much smaller length scales (and therefore at a much higher temperature and amplitude) to exponentially larger lengths, so that they can become  the size of the Hubble radius today. In other words, the entropy production at the end of inflation  makes the ``adjacent'' contraction and expansion phase sufficiently asymmetric to ``amplify'' the thermal fluctuations.

({\bf iii}) Consequently, one only needs a power-law inflationary phase, $a\sim t^p$, with $p> 1.3$, a requirement that can be satisfied with exponential potentials naturally arising in supergravity/string theory compactifications~\cite{exp-pot,flux,bj,munoz,bbc}. ({\bf ii}) A second rather unique feature of this model is that the number of e-foldings in the inflationary phase actually controls the amplitude of fluctuations, and the observed value in the sky automatically guarantees that this model does not suffer from the  transplanckian problem. We shall see that the fluctuations are produced when both classical thermodynamics and General Relativity are expected to be valid. ({\bf iv}) Further, in this model the inflationary phase is preceded by a phase of thermal matter which in some sense prepares the initial condition for inflation to begin.  Although, whether this can really address the ``low entropy'' problem,  is not clear. Finally, by embedding inflation in the ``emergent cyclic universe'' scenario we will be able to address {\bf (i)} concretely.

While cyclic models are employed to describe past-eternal non-singular cosmologies, it is remarkably difficult to achieve this ``in practice''.  Firstly, one needs to come up with a theoretically consistent (ghost and instability free) new physics to resolve the Big Bang singularity. In General Relativity there are strict ``no-go'' theorems prohibiting  ``bounces''~\cite{paris-ec} where contraction gives way to expansion. Fortunately encouraging progress have been made in the recent past, mostly using non-local and/or non-perturbative physics, such as in Loop Quantum Cosmology\footnote{In  Brane-world scenarios with extra time-like directions~\cite{sahni}, one also has an ``effective'' negative energy contribution which goes like $\rho^2$ similar to what is found in Loop Quantum cosmology which can therefore also lead to singularity free bouncing universes.}~\cite{loop}, string inspired non-local modifications of gravity~\cite{bms},  stringy toy models using AdS/CFT ideas~\cite{turok}, tachyon dynamics~\cite{tachyon}, and mechanisms involving ghost condensation~\cite{ghost}, and fermion condensations (both classically~\cite{greene} and via quantum  BCS-like gap formation~\cite{ab}).

Secondly, even if we are able to avoid the BB singularity, in cyclic models we are confronted with the second law of thermodynamics according to which the total entropy in the universe can only increase monotonically. As first pointed out by Tolman~\cite{tolman}, this immediately tells us that the evolution can at most be ``quasi-periodic'' where the length of the cycle typically  monotonically increases with the increase in entropy from cycle to cycle (see~\cite{columbia} for a counter-example). Moreover,  the problem of the ``beginning of time'' comes back to haunt us, as  the cycles  either become vanishingly small at a finite proper time  in the past~\cite{tolman} or is geodesically incomplete, as in the past-eternal inflationary scenarios~\cite{vilenkin}.

\begin{center}
\begin{figure}
\includegraphics[height=6cm,angle=0,scale=1.5]{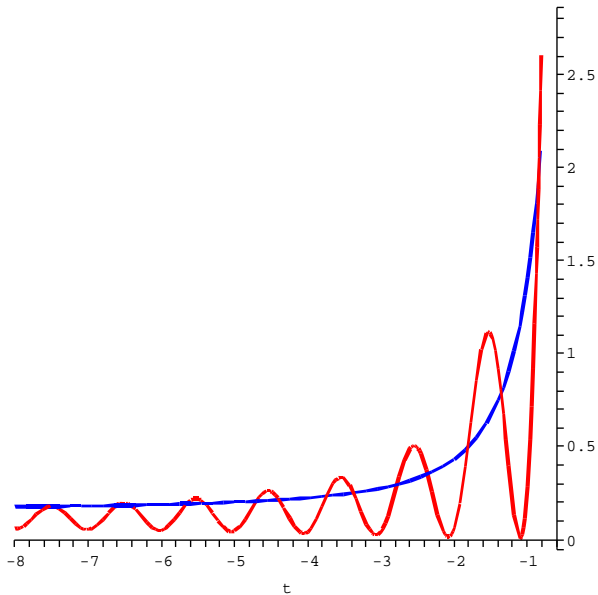}
\caption{ \label{figemergent} Qualitative plot of how the scale factor evolves as a function of the conformal time $\tau$. The cycles have the same period in $\tau$, but they grow as a function of proper time. The blue curve denotes the transition between the Hagedorn (below the curve) and the non-thermal (above the curve) phase. The evolution ends in an inflating expanding branch. }
\end{figure}
\end{center}

In~\cite{emergent} a new ``emergent'' cyclic universe model was presented where it was argued that String theory may have the ingredients to cure both the Big Bang and Tolman's Entropy problem. Repulsive Casimir energies coming from scalar massless string states were invoked to obtain the bounce. (A more detailed accounting of all other different contributions to the Casimir energy is  currently under study~\cite{radu}.)  In~\cite{emergent} it was also argued how the existence of a thermal Hagedorn phase in string theory at high enough energies/temperatures can  provide a nice solution to Tolman's entropy problem. The important implications of the Hagedorn phase in String-Gas-Cosmology~\cite{vafa} (see~\cite{string-gas-review} for recent reviews) has been discussed several times~\cite{tseytlin,string-thermo,anupam,hindmarsh} in the past, but only recently in the context of cyclic models~\cite{emergent,columbia}. In particular in~\cite{emergent} it was shown how for a large class of closed universe models, as we go back in the past (cycles) the universe starts spending more and more time in the thermal Hagedorn phase where entropy remains constant. As a result, in the infinite past the universe assymptotes to an almost periodic  evolution with short time periods (string time scale and energy densities) and constant but non-zero entropy, see fig.~\ref{figemergent}.

In each of these cycles some amount of entropy is always produced in the non-thermal phase (gas of radiation + non-relativistic matter) via exchange of energy from matter to radiation and as a result the cycles and the universe gradually becomes longer and larger respectively~\cite{emergent}. Effectively, because of entropy production in every cycle the universe expands a little more than it contracts and this process can continue forever.

 In this paper we explore the much richer dynamics that can emerge once we include scalar fields, and in particular with exponential potentials~\cite{exp-pot,flux,bj,munoz,bbc} and couplings~\cite{vafa,tseytlin,bbc,string-gas} to matter, which are quite common in string and supergravity theories. Depending upon the exponents, we will find that there can now be two different phases: In the initial ``emergent'' phase the scalar field is essentially a spectator, the cycles grow gradually and the evolution is qualitatively similar to what  was discussed in~\cite{emergent}. There is no inflationary phase in these cycles. However, when the entropy of the universe reaches a critical value, the scalar field suddenly becomes important and  starts to dominate the energy density  of the universe. Further, if the scalar field potential is flat  enough, then one enters a phase of  inflation which prevents the spatial curvature driven turn-around, making it the last  cycle!  It is already well known that the exponential potentials that one obtains in stringy compactifications can indeed give rise to a ``power-law'' phase of inflation, especially in the context of string gas cosmology~\cite{bbem,bbmm}, but typically with a very small power which is unable to reproduce the spectral tilt observed in CMB. Fortunately as mentioned before, in our scenario we do not require inflation to produce the scale invariant perturbations.

Finally,  our scenario can be completely consistent with the recent observation of dark energy by  including the usual tiny cosmological constant.

In the next section we will start with a review of the emergent cyclic universe model that was presented in~\cite{emergent}, and discuss the evolution in the early short cycles. Next in section \ref{spectrum}, we will discuss how in this cyclic universe model one can incorporate a ``new'' mechanism for producing the fluctuations that we see in our sky today. In the subsequent section \ref{toy}, we will provide a specific toy model realization of the above mechanism via a phase of power-law inflation in the last cycle.   Finally in section \ref{conclusions}, we conclude with an outlook towards future direction.

%%%%%%%%%%%%%%%%%%%%
\section{Review of Emergent Cyclic Models}\label{early}
 In String theory there appears a rather curious UV phase where all the string states (massless and massive) are in thermal equilibrium and therefore entropy is conserved. This allows one to circumvent Tolman's entropy problem by constructing an ``Emergent cyclic universe''~\cite{emergent} where as one goes back in cycles (time) and the cycles become shorter and on an average hotter, it spends more and more time in the Hagedorn phase where no entropy is produced. Thus the universe asymptotes to a constant entropy state with almost-periodic contractions and expansions. In these cycles entropy is produced only below a certain ``critical phase transition temperature'', $T_p$, when the different species in the Hagedorn phase fall out of thermal equilibrium and energy starts to flow from hotter to colder species. (A most common example of this kind that we observe today, which incidentally inspired Tolman, is that of radiation being emitted from ``hot'' stars in space containing the ``cold'' background radiation.) As a result we find two qualitatively different possible cosmological histories depending on the ``initial data'': If the asymptotic value of entropy at $t=-\infty$, let us call it $S_{-\infty}$, is less than some critical value, $S_{cr}$ (which depends on the ``initial'' volume), then the universe always remains confined in the Hagedorn phase. No entropy is ever produced resulting in a phenomenologically uninteresting periodic evolution. On the other hand, there is a second class of solutions where the lengths of the cycles increase monotonically with time. Each of these cycles spend some amount of time in both the Hagedorn and non-thermal phases, and as a result some amount of entropy is always produced.  However, as one goes back in the past the universe spends less and less time in the entropy producing non-thermal phase, the entropy produced in a given cycle, $\De S\ra 0$, and  the universe asymptotes to a periodic evolution with $S_{-\infty}=S_{cr}$ thereby avoiding Tolman's problem.

This new cosmological scenario also naturally explains the flatness  and horizon problems. The universe can ``start'' out with short cycles where  energy densities in spatial curvature (which also gives us a turn-around  in our model) and matter are comparable,  near the string scale. Yet through the course of entropy production  in the infinite past cycles, the evolution eventually produces cycles with much larger entropies so that  curvature  can remain negligible for a very long time.  Through the course of these infinite number of cycles there is also obviously enough time to establish causal connection.  The only {\it a priori} assumption that we have to make is that of  spatial homogeneity/isotropy. Thus an  important open question  is whether one can find a mechanism which can drive a generic background (or at least a patch) to a homogeneous and isotropic space time?

 One of the key difference between the approach advocated in~\cite{emergent} and here,  to some of the previous ones is that we incorporate  non-singular bounces by invoking negative Casimir energy.  This means that  the Hubble rate remains  bounded and around  $\sim T_H^2/M_p$ where $T_H$, the Hagedorn temperature  is close to the String scale, and $M_p$ is the reduced Planck scale. Now we expect a thermal  Hagedorn phase to exist provided~\cite{anupam,hindmarsh}
\be
{\Ga_{sc}\over H} \approx {ng_s^2T_H\over H_{\mtx}}>1
\ee
where $\Ga_{sc}$ is the  scattering interaction rate that keeps the various string states in thermal equilibrium, $g_s$ is the string coupling, and $n$ is the number of strings present per unit string volume.  To maintain thermal equilibrium we only need to satisfy $g_s^2>T_H/nM_p$, a rather modest requirement.  If $T_H\ll M_p$, and the string coupling is not too small, thermal equilibrium can easily be maintained, and hence forth we are going to assume that we indeed have a thermal Hagedorn phase around the bounce. In a singular big crunch/bang scenario the interaction rates which can potentially maintain thermal equilibrium among the different  species,  cannot possibly keep up with the {\it diverging}  Hubble rate.   Thus most previous literature  assumes that such transitions will produce large amounts of entropy during the crunch-bang transition, which eventually leads to Tolman's  problem.

Let us now try to understand how the emergent cyclic behavior is realized in more details.
\subsection{Hagedorn Physics, Casimir Energy and Bounces: }
We start by considering the Hagedorn phase of matter on a space-time which includes our three dimensional spatially closed universe, $S^3$, along with appropriately compactified extra dimensions\footnote{One may be worried about consistency of string theory on curved manifold such as the three sphere, $S_3$. Although little known string theory can be consistent in ``curved backgrounds''~\cite{curved}. In particular, it was shown in~\cite{peter}, that for  group manifolds anomalies can cancel provided the Casimir invariants of the group satisfy certain algebraic relations. In particular, it is easy to check that $SO(3)\times SO(3)\times SO(3)\approx S_3\times S_3\times S_3$ is such a consistent spatial background. }. Since $S^3$ does not have any non-trivial homotopy,  strings do not possess any winding number, and the thermodynamics of the string-gas should be the same as for three non-compact directions~\cite{jain}, but in any case,  we will only care about the ``universal'' leading order behavior of entropy~\cite{jain,vafa}
\be
S=\bb_H E+\cO(VT_H^4/E)\Ra \rho_{\mt{hag}}\approx {T_H^4S/V}
\label{hag-entropy}
\ee
where $S$ and $E$ are the entropy and energy of the Hagedorn phase, $V$ is the spatial volume of the universe, and $T_H$ is the Hagedorn temperature. The most important aspect of Hagedorn physics is that all the different string states are in thermal equilibrium with each other and the entropy of the system remains constant. This is the crucial property which will help us resolve Tolman's problem of ever-shrinking cycles.

We will now work in a closed universe model because it is known that the periodic boundary condition yields repulsive negative Casimir energy contributions. For the purpose of illustration we only consider here the Casimir energies coming from minimally coupled massless scalar fields~\cite{casimir}. In string theory massless moduli scalar fields are ubiquitous, but typically they also couple to the Ricci scalar. Such couplings may modify the form of the Casimir energy. In principle one has to also include contributions to Casimir energies coming from all the massive string states, not just the massless ones, but we leave these detailed calculations for future~\cite{radu}. As an added bonus, the spatial curvature from $S_3$ can provide us with ``turn-arounds'' when expansions give way to contractions, an essential ingredient toward constructing a cyclic cosmology. The Hubble equation therefore reads as
\be
H^2={T_H^4\over 3 M_p^2}\LT {S\over a^3}-{\Omega_c\over a^4}-{\Omega_k\over a^2}\RT
\label{hag-hubble}
\ee
where $\Om_c,\Om_k$ are constants, and we have defined a dimensionless ``scale factor''
\be
a\equiv T_H V^{1/3}
\ee
The first term corresponds to energy density of the Hagedorn phase. The second comes from the negative Casimir energy contributions~\cite{casimir}
\be
\rho_c=-{\Omega_cT_H^4\over a^4}\ ,
\label{casimir}
\ee
while the third term reflects the fact that we are in a closed universe.

One can solve the Hubble equation (\ref{hag-hubble}). Explicitly one finds
\be
a(\tau) = {S-\sqrt{S^2-4\Om_c\Om_k}\cos \nu\tau\over 2\Om_k}\mx{ with } \nu\equiv \sqrt{\Om_k\over 3}{T_H^2\over M_p}
\label{periodic}
\ee
where $\tau$ is the usual conformal time defined via the following gauge-choice of the FLRW metric:
\be
ds_4^2=a^2(\tau)\LT-d\tau^2+(1-\Om_kT_H^4r^2/3M_p^2)^{-1}dr^2+r^2d\Om^2\RT.
\ee
 If  both the bounce and the turn-around occurs within the Hagedorn phase then we end up having an ``eternally periodic cyclic universe'' given by (\ref{periodic}).

 Clearly, this describes a non-singular universe but a rather uninteresting one, the temperature  cannot fall much below the Hagedorn temperature as then the string gas will no longer remain in the Hagedorn phase, the assumption under which (\ref{periodic}) was derived. As we will see later, the  periodic solution  holds as long as the entropy is less than some critical value.
Our primary interest lies in exactly the situation when  (\ref{periodic}) does not hold, and the cycles can grow with the production of entropy in the ``non-thermal'' phase.

%%%%%%%%%%%%%%%%%%%%%%%%%%%%%%%%%%%%%%%%%
\subsection{Ordinary matter, Curvature and turn-arounds}
From simple thermodynamic considerations it follows that once the different species fall out of equilibrium and there is exchange of energy (from the hotter and entropically less favorable to colder entropically favorable species), entropy in generated.  We want to incorporate such entropy production  in a toy model setting. For concreteness we will consider a two species model, radiation, $\rho_r$, and dust-like matter,  $\rho_m$, and assume that the entire string gas can be clubbed into  two of these categories near the transition from Hagedorn phase to radiation. Thus the picture is, that both species are in thermal equilibrium in the Hagedorn phase, but below the transition temperature the two species fall out of equilibrium and energy starts to flow from the dust-like matter to radiative degrees of freedom.

Before we model this energy exchange, it is instructive to look at the usual solution of a closed universe with ordinary matter and radiation. The Hubble equation is of the same form as (\ref{hag-hubble}) with $S\ra\Om_m$ and $\Om_c\ra \Om_c-\Om_r$:
\be
H^2={\rho_m+\rho_r-\rho_c-\rho_k\over 3 M_p^2}\equiv{T_H^4\over 3 M_p^2}\LT {\Om_m\over a^3}+{\Omega_r-\Om_c\over a^4}-{\Omega_k\over a^2}\RT
\label{2nd-hubble}
\ee

The solution is also of the same form as (\ref{periodic})
\be
a(\ti{\tau}) = [\Om_m+\sqrt{\Om_m^2+4(\Om_r-\Om_c)\Om_k}\cos \nu\ti{\tau}]/ 2\Om_k
\label{scalefactor}
\ee
 Above, $\ti{\tau}$  again denotes the conformal time, but in a different patch surrounding the turn-around, $\ti{\tau}=0$.

To complete the story one needs to relate the quantities in the two halves of the cycle.  To keep the calculations simple, we are  going to assume that one can ignore $\Om_c$ as compared to $\Om_r$ which is possible provided $\Om_c\ll\Om_k^2$, see appendix \ref{omc}. Unfortunately near the transition, the thermodynamics of the Hagedorn phase is not well understood as  the corrections in (\ref{hag-entropy}) around Hagedorn energy densities are not under control, but by our explicit construction the total entropy should be conserved above the transition temperature. Now, the entropies in matter and radiation are given by
\be
S_r={4\rho V\over 3T}={4\over 3}g^{1/4} \Om_r^{3/4}\mx{ and } S_m={\rho V\over M} ={T_H\over M}\Om_m\equiv b_m\Om_m
\label{entropy-formula}
\ee
where $M\sim T_H$ corresponds to the mass of the non-relativistic particles and in our convention,  $g=\pi^2/30$ times  the number of ``effective'' massless degrees of freedom. Thus in a two-species toy model and in the approximation of an instantaneous transition\footnote{The transition obviously is going to be much more complicated in reality as many string species will be involved with different scattering cross-sections, and a much more in depth study will be required to model the dynamics accurately,  but this is out of the scope of the present paper. We note that in~\cite{string-thermo,columbia}   such  analysis have been performed when the ten dimensional manifold is a tori. This shows that in principle such studies can be done, at least within some approximation schemes.} we have the first matching condition
\be
S= {4\over 3}g^{1/4}\Om_r^{3/4}+b_m \Om_m
\label{total-entropy}
\ee
We also know that at the point of phase transition  matter and radiation were still just in thermal equilibrium, \ie had the same temperature
\be
T_m={T_H\over b_m}={T_H\Om_{r}^{1/4}\over  ag^{1/4}}=T_r
\label{temperature}
\ee

Finally, we make a phenomenological ansatz about the relative energy densities of matter and radiation at the transition epoch
\be
\mu\equiv \rho_{m}/\rho_r=\Om_{m}a_p/\Om_r
\label{mu}
\ee
Ideally, this ratio should be calculable if we understood the transition from Hagedorn to radiation+matter phase. If we want this matter component to be mostly stable and ultimately correspond to the dark matter particles, then phenomenological considerations (matter-radiation equality  $\sim 100\ ev$) suggests $\mu$ to be very small $\sim 10^{-22}$, see appendix \ref{muvalue} for details. To understand the basic physics of how the emergent cyclic scenario works we will assume this for now. In the next section, we will consider a slightly different and perhaps more realistic scenario where after the transition the universe will be dominated  by the ``relic'' matter density and significant radiation is only going to be produced when these massive states decay. The underlying physics of how we solve Tolman's entropy problem will however remain unchanged.

Using (\ref{total-entropy}), (\ref{temperature}) and (\ref{mu}), we can now determine all the dynamical quantities involved in the second half in terms of the phenomenological constants, $b_m, g, \mu$ and the entropy of the Hagedorn phase (which varies from cycle to cycle):
\bea
\Om_r&=&\LT{S\over b_r(1+3\mu/4)}\RT^{4/3}\mx{ , }\Om_m={3\over4}\LT{\mu S\over b_m(1+3\mu/4)}\RT\nonumber\\
a_p&=&{4b_mS^{1/3}\over 3b_r^{4/3}(1+3\mu/4)^{1/3}}
\label{gen-matching}
\eea
 Under the simplifying assumptions $b_m\sim 1\Leftrightarrow T_p\sim T_H$ and $\mu,\Om_c/\Om_k^2\ll 1$ the expressions simplify to
\be
\Om_r={1\over g^{1/3}}\LF{3S\over 4}\RF^{4/3}\ ,\ \Om_m={3\over4}\mu S\mx{ and }a_p=\LF{3S\over 4g}\RF^{1/3}
\label{matching}
\ee

%%%%%%%%%%%%%%%%%%%%%%%%%%%%%%%%%%%%%%%%%
\subsection{Entropy Production}
Having understood the behavior of our universe in a ``non-interacting'' entropy-preserving setting, let us now try to understand how energy exchange between ordinary matter and radiation effects the dynamics.
The energy exchange process can be captured via phenomenological equations of the form~\cite{barrow-exchange,barrow}
\be
\dot{\rho_r}+4H\rho_r=T_H^4 s\mx{ and }\dot{\rho_m}+3H\rho_m=-T_H^4s
\label{continuity}
\ee
where $s$ characterizes the energy flow and typically one expects it to depend on the densities and the scale factor. A number of phenomenologically interesting cases can be captured by equations of the above form, see for instance~\cite{barrow-exchange}. For future reference we note an especially important case for us, that of matter decaying into radiation for which $s\propto \Ga \rho_m/T_H^4$, where $\Ga$ corresponds to the usual decay rate. It is easy to compute the net entropy production in such models:
\be
\dot{S}=\dot{S_r}+\dot{S_m}=a^3s\LF{g^{1/4}T_H\over \rho_r^{1/4}}-b_m\RF=a^3s\LF{T_H\over T_r}-{T_H\over T_m}\RF
\label{entropy-growth}
\ee
$s>0$, corresponds to energy flowing from matter into radiation. Consistency with $2^{nd}$ law of thermodynamics then implies that the quantity within brackets must be positive. This is nothing but the condition that the temperature of the non-relativistic gas be greater than the radiative gas, so that energy flows from the hotter non-relativistic species to colder radiation in accordance with $1^{st}$ law of thermodynamics. Since in our picture the two species have the same temperature $T_p$, at the transition point, where after $T_r$ decreases, while $T_m$ stays fixed, (\ref{entropy-growth}) is consistent with both the $1^{st}$ and  $2^{nd}$ law of thermodynamics. We also note in passing that the modified continuity equations (\ref{continuity}) obviously satisfies conservation of the total stress energy tensor. However, the energy exchange term breaks the time-reversal symmetry  providing the arrow of time in the direction of increasing entropy.

Before performing a quantitative estimate of the entropy growth with cycles, let us physically see why we can realize the emergent cyclic universe in the first place. For this purpose it is sufficient to look at short cycles which (we claim) asymptotically approaches the periodic evolution. In these cycles matter density is always negligible as compared to radiation, see the discussion around (\ref{mu}) and the appendix \ref{muvalue}, so that the expression for the turn-around point reduces to
\be
a_{\mtx}\approx {\sqrt{\Om_r/\Om_k}}\approx S^{2/3}/(  \sqrt{\Om_k}b_r^{2/3})
\ee
We now come to a crucial point in the paper. We notice that while $a_{\mtx}\sim S^{2/3}$, $a_p\sim S^{1/3}$. In other words as we go back in the past and the entropy decreases, $a_{\mtx}$ catches up with $a_p$, and the universe spends less and less time in this entropy-generating phase. In turn, less and less entropy is produced,  and in fact the entropy approaches a constant given by
\be
a_{\mtx}=a_p\Ra \lim_{n\ra -\infty}S_n= S_{cr}\equiv (4/ 3)^3\Om_k^{3/2}b_r^{2}
\label{critical}
\ee
(\ref{critical}) also provides us with a ``boundary'' between two qualitatively different universes: {\it (A)} eternally periodic cyclic universes given by (\ref{periodic}), where no entropy is produced and $S<S_{cr}$. {\it (B)} emergent cyclic universes, where the entropy grows   with cycles in the future direction, while assymptoting to a constant, $S_{cr}$ in the infinite past. In the (B) type evolution this ``initial'' entropy, $S_{-\infty}(=S_{cr})$, or equivalently the ``initial volume'' $a_{p,-\infty}$, is a free parameter which is encoded in the value of $\Om_k$.

We will now try to estimate  the entropy increase in a given cycle, when the cycles are ``short'' \ie, $\nu\taut_p\ll 1$, where $\taut_p$ is the conformal time corresponding to the transition point $a_p=a(\taut_p)$.
Now, since $\taut_p\ra 0$, to capture  the leading behavior in entropy growth, we can approximate $s$ to be a  constant, $s=s_{cr}$, where $s_{cr}$ is evaluated by substituting  $a=a_{\mt{max}}=a_p$  in $s=s(a,\rho_m(a),\rho_r(a))$, as appropriate for cycles approaching the critical one.  Assuming that the decay rate is small as compared to the the time period (technically this means $s\ll T_H^2/M_p$), it is sufficient to keep only linear order corrections in $s_{cr}$ in (\ref{entropy-growth}).  In this case in (\ref{entropy-growth}) one can ignore the $s$-dependence on $\rho_r, a(\tau)$, and  use all the formulas derived  previously in the ``no energy exchange'' limit.

One thus has
\be
{dS\over d\tau}=a^3s\LF{a\over a_p}-1\RF
\label{entropy-dot}
\ee
Then using (\ref{scalefactor}) and (\ref{entropy-dot})  one straight-forwardly obtains
\be
\Delta S={2a_{\mtx}^4s_{cr}\nu^2\taut_p^3\over 3a_p}\approx {d\de S\over dn}
\label{deltaS}
\ee
where ``$n$'' labels the number of the cycle and $\de S\equiv S-S_{cr}$.
 Since, $a_{\mtx}, a_p$ and $\taut_p$ are all known functions of the entropy, this differential equation can be solved.
As $n\ra -\infty$ we find the leading order result:
\be
S\approx S_{cr}\LT1+{1\over C^2 n^2}\RT\mx{ where } C= 8\LF{2\over 3}\RF^{11/2}{s_{cr}\over \nu b_r^4}
\ee
We now explicitly see that, $S\ra S_{cr}$ at the past infinity.
%%%%%%%%%%%%%%%%%%%%%%%%%%%%%%%%%%%%%%%%%
\section{Generating CMB Fluctuations}\label{spectrum}
\subsection{Spectrum}
Let us now turn our attention to one of the key issues in observational cosmology, that of generating near scale invariant CMB fluctuations. Several different alternatives have been considered in the literature with varying degrees of success. In ekpyrotic models one considers scalar field fluctuations, and while single field models do not typically give rise to a scale-invariant spectrum~\cite{ekpflucts} (but see~\cite{abb} for a possible exception) the addition of other scalar fields can cure this problem via curvaton mechanism~\cite{justin-new}. In~\cite{nayeri} instead, thermal stringy fluctuations in the Hagedorn phase were considered and in the context of   a specific bouncing universe scenario they were shown to give rise to a scale-invariant spectrum~\cite{bbms}, addressing some of the concerns raised in~\cite{linde-hag}. More recently new mechanisms involving different kinds of matter in the contraction phase have been proposed~\cite{matter}. In this paper  we consider thermal fluctuations that are generated in the General Relativistic (\ie temperature much below the Planck scale) radiation dominated  contracting phase just preceding the  phase of expansion ``we'' are in.

Some time back Peebles had observed that in general if thermal fluctuations in  radiation can be  ``frozen-in'' at some earlier higher temperature, it can potentially explain the observed CMB fluctuations~\cite{peebles}. As noted in~\cite{pogosian} a first requirement to make this work is to have a radiation dominated contraction phase which naturally arises in our model.   To see in some detail let us start by  noting that the amplitude of fluctuation, $\de_L$,  in energy in a typical volume of size $L^3$ in a thermal environment can be computed from the partition function, $Z(V,\bb\equiv T^{-1})$:
\be
\de^2_L={d^2ln Z\over d\bb^2}/\LF{d^2ln Z\over d\bb^2}\RF^2={T^2\over \rho^2 L^3}{\p \rho\over \p T}
\label{deltaL}
\ee
where in the last step we have assumed that energy is an extensive quantity and goes like the volume, \ie the energy density only depends on temperature. Using appropriate window functions one  then finds a white noise power spectrum~\cite{pogosian}
\be
\de^2_L={4\over g T^3L^3}\Ra P(k)=<\de_k^2>={32\pi^{3/2}\over g T^3}\approx {\cO(100)\over gT^3}
\label{power-spectrum}
\ee
The crucial point to note however is that the above result is valid only as long as the length scales in question are sub-Hubble and thermal correlations can be maintained, which is a much stronger requirement than just having local thermal equilibrium. Afterwards, in the super-Hubble phase the evolution of density fluctuations is governed by the gravitational potential, $\Phi_k$,  according to the usual General Relativistic perturbation equation
\be
k^2\Phi_k\approx a^2\rho\de_k\sim a^2H^2\de_k
\label{newtonian}
\ee
Now in a radiation dominated contraction phase modes do indeed exit the Hubble radius. At the Hubble crossing we have $ k\sim Ha\sim T$ which  straight away gives us the desired scale-invariance of $P(k)$, \ie $P(k)\sim k^{-3}$. Moreover, along with (\ref{newtonian}) it also ensures that $\Phi_k\sim \de_k$ at the Hubble crossing, and the $k$-dependence is transferred to $\Phi_k$ which then remains approximately constant till the modes again re-enter the Hubble radius in the expanding branch. Thus as long as the modes that we see today exit the Hubble radius in the contracting phase ``sufficiently''far away from the turnaround and the bounce, where the spatial curvature and Casimir energy can become important, $\Phi_k$ will obtain a scale invariant spectrum. Therefore assuming that the  modes don't mix and  $\Phi_k$ is approximately constant during the bounce,  the scale-invariance of thermal fluctuations turns out to be a   generic result.
%%%%%%%%%%%%%%%%%%%%%%%%%%%%%
\subsection{Entropy production and Amplitude Amplification}
The problem with this rather attractive scenario is that the amplitude of these thermal fluctuations turn out to be  too small in usual cyclic/bouncing models\footnote{A different approach to circumventing this problem involves varying the speed of light as advocated in~\cite{pogosian,vsl}.}. This can be illustrated by computing the amplitude of fluctuation which entered the Hubble radius at matter-radiation equality  (this corresponds to a size of around 1 Mpc today) for which we know $\de_{\mtc}\sim 10^{-4}$ at the equality  epoch. Say $T_{\widetilde{\mtc}}$ and $a_{\mte}$ are the temperature and scale factor when the same comoving distance exited the Hubble radius in the contracting phase. For a symmetric bounce, we  have  $T_{\mte}=T_{\mtc}$, so that  according to (\ref{deltaL})
 \be
 \de_{\mte}^2={4\over g_c(T_{\mtc}L)^3}\sim\sqrt{g_c}\LF{T_{\mtc}\over M_p}\RF^3\sim 10^{-72}
 \ee
 However, the production of entropy and the resulting asymmetry between the expanding and the contracting branches in our model  provides a  remarkable opportunity to fix this problem. Let us first estimate the amount of entropy that needs to be produced to obtain the required amplitude. Using the Hubble crossing conditions when the modes exit and then re-enter the Hubble radius in the contracting and expanding phases respectively, and the approximate constancy of $\Phi_k$ in between, it is clear that $\de_{\mtc}=\de_{\mte}$ so that
 \be
10^{-8}\approx \de_{\mtc}^2={4\over g_{\mte} T_{\mte}^3L^3}\sim\sqrt{g_{\mte}}\LF{T_{\mte}\over M_p}\RF^3\Ra T_{\mte}\sim {M_p\over 1000 g_{\mte}^{1/ 6}}
\ee
We remind the readers that at such high energy densities $g_{\mte}$ is expected to be very large ensuring $T_{\mte}\ll M_p$, so that we can trust classical General Relativity which therefore justifies our calculations. Now, using the Hubble crossing conditions:
\be
k= \sqrt{g_{\mte}}T_{\mte}^2a_{\mte}=\sqrt{g_c}T_{\mtc}^2a_{\mtc}
\label{hcrossing}
\ee
 and the expression of entropy in terms of the scale factor: $S\sim (aT)^3$, we find that for this mechanism to work we need
\be
{S_0\over S_{-1}}\sim\LF{T_{\mte}\over T_{\mtc}}\RF^3\sqrt{g_{\mte}\over g_{\mtc}}\sim 10^{64}
\label{generation}
\ee
Here 0 and $-1$ labels our current cycle and the one previous to ours. Thus it is clear that if we can find a mechanism to produce a large amount of entropy then we can make this scenario work.

As an aside, let us note that in this scenario  we don't have to worry about the transplanckian problem. From (\ref{generation}) and (\ref{hcrossing}) one finds that
\be
{a_{\mt{eq}}\over a_{\mte}}\sim 10^{42}
\ee
and thus the physical length, $\equiv\ti{H}_{0}^{-1}$, of the fluctuation corresponding to the current Hubble radius, at the time of it's exit during contraction is given by
\be
\ti{H}_{0}^{-1}=H_0^{-1}10^{-42}\sim 10^{20}M_p^{-1}
\ee
Clearly this is much larger than the Planck length.

Let us return to the problem of sufficient entropy generation. As a first attempt, let us see how much of entropy can be produced by decay of some relic massive particles. For this purpose, let us modify the scenario presented in  section \ref{early} slightly, and assume that after the transition from Hagedorn to non-thermal phase we are dominated by some massive ``relic'' string states (instead of radiation). If these decay into radiation at energy density scale $\sim T_d$, where ``$d$'' denotes the decay epoch, then as is well known, we will produce a large amount of entropy. Now, since entropy goes like volume, the ratio of the entropy during the expansion and in  the preceding contraction  phase at the same energy density scale $\sim T_d$, is given by
\be
{S_{n+1}\over S_n}=\LF{a_{d,n+1}\over a_{d,n}}\RF^3={T_p\over T_d}
\label{matter-decay}
\ee
The expansion in the volume occurs because the universe contracts a little less in the radiation dominated era as compared to the expansion in matter era. Since we want to preserve the successes of Big Bang Nucleosynthesis, $T_d>Mev$. From now we are always going to  assume
\be
T_p\sim T_H\sim 10^{16} Gev
\ee
to keep things concrete and not loose the essential physics in the plethora of parameters. Then, according to (\ref{matter-decay})
we can at most produce an entropy growth of
\be
{S_{n+1}\over S_n}=10^{19}
\ee
This is clearly insufficient.

Traditionally it is well known that any kind of phase transition produces large entropies, a glaring example of the same being ``reheating'' after the end of inflation when the energy of the inflaton field is converted into radiation. Can we therefore embed inflation in our model?  It turns out that the framework of string gas cosmology indeed provides such a  natural extension/modification of the model~\cite{bbem,bbmm}. For definiteness, we are going to from now on assume that $g_L^{1/6}\sim 10$, so that $
T_{\mte}\sim 10^{14} Gev$.
This also means that the comoving length corresponding to the current Hubble radius $\sim 3000 Mpc$, the largest scale that we see today, must have exited the Hubble radius during the contraction phase when the temperature was around $T_{\mth}\sim 10^{11} Gev$. Thus to ensure a near scale-invariant spectrum for the fluctuations that we observe, we at least need the radiation dominated contraction phase to last between
\be
10^{11} Gev\sim T_{\mth}<T<T_{\mte}\sim 10^{14} Gev
\ee
%%%%%%%%%%%%%%%%%%%%%%%%%%%%%%%%%%%
\section{Moduli dynamics, Power-law inflation and COBE normalization}\label{toy}
So far in our analysis, the dynamics of all the scalar fields (ie. moduli) that are present in any string theory compactification have been ignored.  We are now going to provide a toy model illustration of how power-law inflationary dynamics involving the moduli may be able to provide us with the entropy growth that we need. We will assume the scalar field potential\footnote{Moduli fields are strictly speaking massless at tree-level, but they typically obtain a non-zero potential from both perturbative and non-pertubative effects.} to be given by a sum of exponentials
\be
V(\phi)=T_H^4[v_1e^{-\al_1\phi}+v_2e^{-\al_2\phi}]\mx{ with }\al_1>\sqrt{2}>\al_2
\label{potential}
\ee
where to keep things simple here we consider only a single such field $\phi$, which can of course be a linear combination of the different moduli fields such as the dilaton, radion, shape, etc. We note that the value of $\al_2$ is so chosen that once it starts to dominate, we have a power law phase of inflation. Now, for the success of the mechanism exponential potentials are not necessary, but technically exponential potentials are easy to study as exact and approximate analytical solutions are well-known. They are also physically well-motivated as they arise from
 fluxes~\cite{flux},  internal spatial curvature and higher dimensional cosmological constant~\cite{bj}, or even from non-perturbative gaugino condensations~\cite{munoz}, also see~\cite{bbc} where it was shown that non-perturbative potentials in KKLT type scenarios~\cite{kklt} behave like exponentials  in some range of the moduli fields. In the Einstein frame the moduli fields typically couple exponentially to stringy matter so that the energy densities in the Hagedorn and the matter phases now looks like~\cite{bbem,bbc,bbmm}
\be
\rho_{h}=e^{\mu_h\phi}T_H^4 {\Om_{h}\over a^3} \mx{ and }\rho_m=T_H^4e^{\mu_m\phi}{\Om_m\over a^3}
\ee
 We recover the ``decoupled limit'' discussed in the earlier section by setting $\phi=0$ and identifying $S=\Om_h$. Now in general, the couplings $\mu$ can vary from species to species, but for simplicity we are  going to assume that $\mu_h=\mu_m\equiv\mu$. Then,  assuming that most of the entropy at the transition goes to matter, we have $\Om_m\approx\Om_h$.

\begin{center}
\begin{figure}
\includegraphics[height=6cm,angle=0,scale=1]{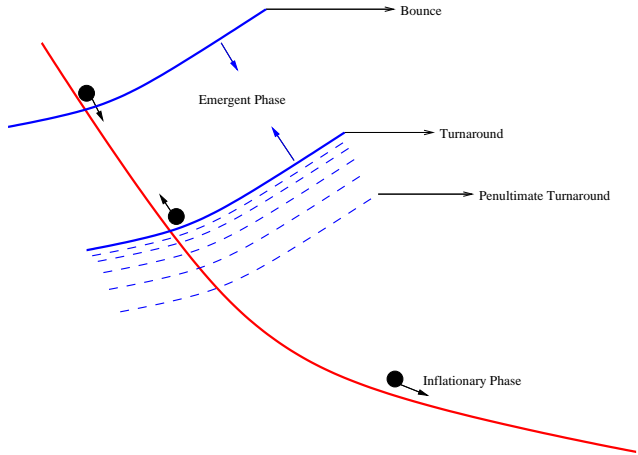}
\caption{ \label{tracking} The red curve corresponds to $V(\phi)$, while the blue curves correspond to $\rho(\phi)$ at the bounce point (the uppermost curve) and turn-around points (the lower curves). The scalar field (represented by the black ball) approximately tracks the minimum formed between  $V(\phi)$ and $\rho(\phi)$. As the entropy and the volume of the universe increases from cycle to cycle,  the turn-around happens at lower and lower energies, until the Penultimate cycle (the lowest blue curve). Before the next turn-around can happen, we enter the inflationary phase which ultimately leads to our universe. }
\end{figure}
\end{center}

 We can understand the entire evolution of the universe as three distinct ``phases'', see fig.~\ref{tracking}: The first phase is the emergent cyclic phase where the cycles are almost periodic, entropy production is small, $S_{n+1}/S_n\lesssim \cO(1)$.  This phase persists till the decay time of the massive string states is much larger than the time period of the cycles. As soon as the two becomes comparable and a significant portion of the matter density can convert into radiation, we have qualitatively a very different cycle, which we call the Penultimate cycle, the reason for such a nomenclature will become clear shortly. Since a large amount of entropy is produced in this cycle, $S_{n+1}/S_n\gg 1$, in the next cycle the universe has to expand a lot more to undergo a turnaround. In the mean time however, the second exponential term in the potential (\ref{potential}) can come into play. Instead of turning around, the universe then enters  a phase of acceleration which makes the universe large and flat, consistent with the universe that we observe today. More importantly it can also produce enough entropy at the end to implement the structure formation scenario discussed in the previous section.   Let us now discuss these phases in details.
\subsection{Tracking Phase, emergent cycles:} This is a phase when the first exponential term dominates the potential, and the scalar energy density tracks the dominant matter component. Approximately the Klein-Gordon equation for the scalar field  reads as
\be
\ddot{\phi}+3H\dot{\phi}=T_H^4\LF\mu e^{\mu\phi}{S\over a^3}+\al_1v_1e^{-\al_1\phi}\RF
\ee
  Such systems have been studied widely in literature in the context of quintessence~\cite{quintessence,bbmm} and inflation~\cite{inflation}, in particular in the context of string gas cosmology~\cite{bbem}, and are known to exhibit tracking behavior, where the scalar field always remains in the minimum of the effective potential $V_{\mt{eff}}(\phi)=\rho(\phi)+ V(\phi)$,  evolving  adiabatically with the red(blue) shifting of the matter species in expanding(contracting) phase:
\bea
 V_{\mt{eff}}(\phi_{min})=T_H^4\LF{\al_1+\mu\over \al_1}\RF\Om_m^{\al_1\over\mu+\al_1}a^{-3\al_1\over\mu+\al_1}
\eea
where  we have chosen the convention $v_1=\mu/\al_1$ to simplify calculations.
The universe evolves as if sourced by an ideal fluid with equation of state
\be
\om=-{\mu\over \mu+\al_1}
\label{omega}
\ee
The condition for exhibiting such tracking behavior is given by
\be
\al_1(\al_1+\mu)>3
\ee
which can very easily be satisfied.

One can easily check that if the transition from Hagedorn to matter phase happens at some constant energy density in each cycle, this also means that it occurs at the same value for $\phi$. Specifically if we say that the transition happens when $\rho_{\mt{hag}}= T_p^4\sim T_H^4$, then one can check that at the transition,  $\phi=0$ (this is why we chose the above convention for $v_1$) and one can just identify $\Om_m=\Om_h=S$, the entropy in the Hagedorn phase.

The dynamics of the scalar field is actually quite simple to visualize. In the Hagedorn phase, the scalar field goes up during the contraction part, $V_{\mt{eff}}$ blue-shifting till it reaches a maximum at the bounce (see fig.~\ref{tracking}), and then as soon as the universe starts to expand, and the Hagedorn matter starts to dilute, $V_{\mt{eff}}(\phi)$ starts to redshift, and the scalar field comes down from where it started at the beginning of the Hagedorn phase. After the transition, in the non-thermal matter phase, roughly the reverse happens. The scalar field energy continues to redshift, now tracking the minimum between the matter component and the first exponential term in the potential (\ref{potential}). However, after the turn-around once the matter density starts to blue-shift, it drags $\phi$ up as well, and the scalar field rises to where it originally started from at the beginning of the Hagedorn phase. Thus the scalar field keeps undergoing oscillatory motion as depicted in the picture fig.~\ref{tracking}.

The evolution of the scalar field is approximately periodic satisfying the approximate Hubble equation
\be
H^2={T_H^4\over 3 M_p^2}\LT \LF{\al_1+\mu\over \al_1}\RF{ S^{1+\om}\over a^{3(1+\om)}}-{\Omega_c\over a^4}-{\Omega_k\over a^2}\RT
\ee
where we have ignored the entropy production in the non-thermal phase and consequently the radiation component. As long as the decay time of the massive particles is much larger than the time period of the cycle, this is a good approximation. We have also ignored the kinetic energy of the scalar field which is known to produce small corrections~\cite{bbmm}, and moreover does not change\footnote{Only near the bounce and the turnaround point the behavior of the coupled fluid will be a little different, and one really needs to perform numerical simulations to determine the exact behavior. This should however not effect the overall picture in any significant way.} the effective equation of state (\ref{omega}).

In order to have the cyclic behavior the spatial curvature must be able to catch up with the coupled fluid. This will happen provided, $\om>-1/3$
\be
\Ra{\mu\over \al_1}<{1\over 2}\,
\label{lowerbound}
\ee
 It is also clear that while as before $a_p\sim S^{1/3}$, $a_T\sim S^{(1+\om)/(1+3\om)}$. As $\om<0$,  the turn-around scale factor will always catch up with the transition scale factor as the entropy decreases in the past giving us the  emergent behavior.

 To summarize, where as in our earlier discussion in section \ref{early}, the Hagedorn phase around the bounce behaved as an $ \om=0$ fluid and the non-thermal phase was dominated by radiation $\om=1/3$, in this coupled system the entire phase is approximately described by a single fluid with an effective equation of state given by (\ref{omega}). The only difference between the Hagedorn and the non-thermal phase is that in the latter matter can decay into radiation thereby producing entropy, as before. As in section \ref{early}, the bounce and the turn-around is still provided by Casimir energy and spatial curvature respectively.
\subsection{The Penultimate Cycle}
In the discussion of our emergent cycles we have assumed that the matter decay is not substantial in the non-thermal phase,  entropy production in each cycle is small and the evolution is therefore very close to being periodic. This is no longer true once the time period (which is roughly the same as the turn-around time) of the cycle becomes comparable to the decay time for the matter species. Since in the emergent phase the entropy, and with it the duration of the cycle increases continually albeit slowly, the above situation inevitably arises, and we will call this cycle the Penultimate one. To be specific, Penultimate cycle occurs  when
\be
\Ga^{-1}\sim \mx{Time Period}\sim {M_p\over T_d^2}\mx{ where }T_d\equiv\rho_{\mt{min}}^{1/4}
\ee
or the entropy becomes
\be
S\sim \LF{T_H^2\over\Ga M_p}\RF^{1+3\om\over 1+\om}
\ee
 After the decay, the universe immediately becomes dominated by radiation. Since the universe is turning around simultaneously, the radiation era  persists as  radiation blue shifts faster than the coupled fluid component.  The mechanism discussed in the previous section to generate scale invariant thermal fluctuations can now be implemented provided $T_d < T_{\mth}$. This ensures that all the scales that we see today will exit during the radiation dominated contraction phase and therefore will be endowed with a scale-invariant spectrum.

Just to emphasize, a crucial difference between the ``emergent cycles'' and the ``penultimate cycle'' is that a lot more entropy is actually generated. Approximately, one finds
\be
{S_{n+1}\over S_n}=\LF{T_p\over T_d}\RF^{1-3\om\over1+\om}
\label{growth-general}
\ee
The above ratio is just the ratio between the volume's\footnote{Since the transitions happen approximately at the same temperature and entropy is proportional to the volume, the ratio of the volume approximately gives us the entropy growth factor.} when the transition from Hagedorn to non-thermal phase and vice-versa occurs during the expanding and contracting phases respectively in the Penultimate cycle. Since $\om<1/3$ during expansion, the universe expands much more in the tracking phase than it contracts during radiation. (\ref{growth-general}) is nothing but a generalization of the well-known case of matter decay (\ref{matter-decay}) and recovers it when $\om=0$.
\subsection{The Last Cycle and Power law Inflation}
We have now seen that once the decay time becomes comparable to the cycle period, a lot of entropy is produced during the decay and the contraction phase is radiation dominated. We can actually calculate what would be the next turn-around point. To get a quantitative intuition, we are going to consider a particular case when $T_d\lesssim T_{\mth}$ and $\mu/\al=3/7\Ra \om=-0.3$.  Now, as before in the Penultimate cycle, the tracking phase will end when the energy density falls to $\sim T_d^4$. However, because this cycle has a lot more entropy than the previous one, the scale factor at the decay is going to be much larger:
\be
{a_{d,0}\over a_{d,-1}}=\LF{S_{0}\over S_{-1}}\RF^{1/3}\approx\LF{T_H\over T_{\mth}}\RF^{1-3\om\over 3(1+\om)}\sim 10^{4.5}
\ee
where the estimated number is for the specific values mentioned above.

Thus the spatial curvature now is much smaller as compared to it's value in the Penultimate cycle, and if we assume a radiation dominated universe after the decay, it takes quite a while longer to turn around. If $k$ labels the curvature, then we have
\be
\left.{\rho_{r,0}\over \rho_{k,0}}\right|_{\mt{decay}}\approx\left.{\rho_{k,-1}\over \rho_{k,0}}\right|_{\mt{decay}}=\LF{T_p\over T_d}\RF^{2(1-3\om)\over 3(1+\om)}\sim 10^{9}
\ee
\be
\Ra {\rho_{\mt{min},0}\over \rho_{\mt{min},-1}}=\LF{T_d\over T_p}\RF^{-4(1-3\om)\over 3(1+\om)}\sim 10^{-18}
\ee
This can be a very small number, as illustrated in the specific example. What this means is that the cycle after the penultimate cycle  lasts a lot longer and the energy density falls a lot more before a turn-around can happen. This gives us a large window to ensure that the second exponential in the scalar field potential comes into play. Since, $\al_2<\sqrt{2}$, the late time attractor solution corresponds to power-law inflation with an effective equation of state
\be
\om_{\phi}=-1+{\al_2^2\over 3}\mx{ and } a(t)\sim t^{2\over\al_2^2}
\ee
 where the energy density quickly becomes dominated by this exponential potential, and the universe cannot turn-around because the spatial curvature cannot catch-up with the inflating scalar field. This phase of inflation can end in the usual way  provided $V(\phi)$ has a minimum, and the universe can then ``reheat''. From then on we will have the usual standard cosmological evolution. Provided the minimum of $V(\phi)\sim (mev)^4$, we will eventually have a phase of dark energy domination consistent with current observation. If this picture is true, we must be living in this eternally lasting last cycle!

Let us now look at the various constraints coming from phenomenological considerations. The first constraint comes from the fact that we need to produce enough entropy at the end of inflation to implement the structure formation scenario discussed in section \ref{spectrum}. Let us say that inflation ends and ``instantaneously'' reheats the universe to a temperature of $T_r$. Then, the net gain of entropy is given by
\be
{S'_{0}\over S_{-1}}\approx\LF{T_p\over T_d}\RF^{1-3\om\over1+\om}\LF{T_d\over T_r}\RF^{1-3\om_{\phi}\over1+\om_{\phi}}
\ee
where the prime denotes the entropy after reheating. In the above estimate we have assumed that the radiation phase after the decay is short \ie,  we enter the inflationary phase rather quickly after the end of the tracking phase. Now, in order to preserve the successes of the big bang nucleosynthesis we must have $T_r>Mev$. On the other hand to obtain the correct COBE normalization for the amplitude of fluctuations we must produce enough entropy (\ref{generation}). As mentioned before, an equivalent way of looking at it is that we need to stretch the fluctuations produced with the correct amplitude at much smaller physical scales to exponentially larger values so that they can correspond to the lengths that we observe in our sky today. This gives us a constraint on $\om_{\phi}$. For the ``canonical'' values that we are considering, we find
\be
\om_{\phi}\leq -0.45\Ra \al_2<1.13
\label{alpha2}
\ee
which is a rather mild requirement to satisfy.
In particular this means a rather low power law inflation is sufficient. This  may be possible to realize in string theory/supergravity~\cite{exp-pot,flux,bj,munoz,bbc}, and specifically in string gas cosmology scenarios~\cite{bbem,bbmm}.
%%%%%%%%%%%%%%%%%%%%%%%%%%%%%%%%%%%%%%%%%%%%%%%
\subsection{The problem of re-entry of modes before inflation}\label{fine-tuning}
To summarize, so far according to our previous discussion all the phenomenological constraints are satisfied provided the parameters of the potential satisfies minimally the following conditions
\be
\al_2<\sqrt{2}\ ,\mu/\al_1< 1/2\ ,\al_1(\al_1+\mu)>3\Ra \al_1>\sqrt{2}
\label{exponents}
\ee
Since the exponents are typically expected to be $\cO(1)$, the above conditions  seem very reasonable to satisfy. We also want the second exponential term not to start dominating before the tracking phase ends through decay. This means that $v_2$ cannot be greater than some critical value, $v_{cr}$. Since in our convention $v_1\sim \cO(1)$, a rough estimate of $v_{cr}$ is given by
\be
v_{cr}\approx \LF{\Ga M_p\over T_H^2}\RF^{\al_1-\al_2\over \al_1}\sim \LF{T_{\mth}\over T_H}\RF^{2(\al_1-\al_2)\over \al_1}
\label{v2}
\ee
For the canonical values, minimally (\ie, assuming $\al_1$ just satisfies (\ref{exponents}), and $\al_2=1.13$  (\ref{alpha2})) we require a mild hierarchy $v_2<v_{cr}\sim\cO(10^{-2})$. Again, this doesn't seem very unreasonable either.

Unfortunately there is another important issue that we have not addressed so far. This issue was first discussed in the context of a two stage inflationary model where there was a break between an exponential and a power-law phase of inflation~\cite{multiple}. While a short-lived exponential phase was required to generate the near-scale invariant fluctuations, the power law phase was needed to solve the usual cosmological puzzles. Our scenario is actually a little similar where  the role of the exponential inflationary phase is played by the radiation dominated contraction era. Now, it was realized in~\cite{multiple} that during the non-inflationary phase modes can re-enter the Hubble radius thereby spoiling their scale-invariant spectrum. We also have to ensure that this does not happen during the tracking and radiation dominated phase in ``our'' expanding branch.

It is now clear that if $v_2$ is much smaller than the critical value then   we will have a long phase of radiation after the decay and therefore run into the above problem. On the other hand,  $v_2$ cannot be too large because then (\ref{v2}) cannot be satisfied, the phase of inflation will start before the particles can decay and  we will not get a radiation dominated contraction phase in the Penultimate cycle.

How close does $v_2$ have to be to the critical value (\ref{v2}) depends on the various parameters of course, but for our canonical choice, it depends most sensitively on $\om$. What we must make sure is that once the mode corresponding to the shortest scale that we observe today (roughly $Mpc$ scale) crosses the horizon during contraction
\be
k_{\mte}=aH\equiv \cH\ ,
\ee
it does not re-enter the horizon before the power-law inflationary phase commences. Minimally this requires that the mode does not enter the horizon in the tracking phase. Since during the radiation dominated contraction phase $\cH\sim 1/a$, while in the tracking phase $\cH\sim a^{-(1+3\om)/2}$ we find
\be
{\cH_d\over\cH_{\mte}}={T_p\over T_{\mte}}\LF{T_d\over T_p}\RF^{2(1+3\om)\over 3(1+\om)}
\ee
We need this ratio to be greater than 1. For our canonical example, $T_p\sim T_H\sim 10^{16}Gev$, and $T_{\mte}\sim 10^{14}Gev$, assuming that the decay is almost immediately followed by inflation, this  gives us the following upper bound on $\om$
\be
\om<-1/ 6\approx -0.17
\label{upperbound}
\ee
Conversely, if $\om$ is less than the above constraint, $v_2$ can be smaller than the critical value, but clearly some fine-tuning will be required. A more comprehensive study  of the fine-tuning required will need numerical exploration which is beyond the scope of this paper.

For the sake of completeness we note that the upper bound (\ref{upperbound}) along with (\ref{lowerbound}) gives us the following constraint on $\mu/\al_1$:
\be
{1\over 5}<{\mu\over \al_1}<{1\over 2}
\ee
Again, this seems a very reasonable range for the exponents to satisfy.
%%%%%%%%%%%%%%%%%%%%%%%%%%%%%%%%%%%%%%%%%
\section{Conclusions}\label{conclusions}
The aim of this paper was to present a qualitatively new phenomenologically viable model of inflation which has some advantages over the conventional slow-roll counterpart. In particular we were able to provide a non-singular completion to inflation in the infinite past, circumvent the trans-Planckian problem, and  most crucially be consistent with CMB with just a low power-law inflationary phase. This was achieved by embedding a string gas inspired power-law inflationary phase in the ``emergent cyclic universe'' scenario which was recently proposed to solve Tolman's entropy problem. The analysis was most certainly done at a toy model level, but was inspired by several ingredients in string theory, such as the existence of a thermal Hagedorn phase in the UV, the presence of  moduli fields  which can give rise to repulsive Casimir energies and the fact that often some of the scalar moduli fields have exponential potentials and couplings to matter. This gives one hope that one may be able to realize this  scenario, or slight variants of it, in a more concrete string theory framework.

In order to achieve this we need in future to examine different aspects of equilibrium and non-equilibrium string thermodynamics as well as include specific moduli dynamics/stabilization. In principle this should also explain how the observed three dimensions became large as compared to the extra dimensions. It would be interesting to revisit the BV mechanism~\cite{vafa} proposed in this context in the light of the new cyclic cosmology discussed here. Another important ingredient in string theory is the existence of higher dimensional objects such as branes, which may in fact play an important role  in explaining the  hierarchy in the sizes between the large observed   and the smaller internal manifolds if they can wrap around the latter~\cite{branes}. We were not able to address how we start from an initial homogeneous and isotropic universe. If however ``initially'' (at $t=-\infty$) the universe has a small anisotropy then in the course of subsequent evolution this anisotropy is only going to get further diluted as the universe, on the whole,  keeps expanding due to entropy production. So, at least if the ``initial conditions'' are fine tuned one should be able to avoid a mixmaster like chaotic evolution. Intriguingly, there are suggestions that a brane gas phase can naturally lead to isotropization of the different directions~\cite{brane-isotropy}, which could essentially predate the emergent cyclic behavior discussed here. There are certainly indications that one may  have new thermal phases involving brane states which can later on pave way to the more conventional thermal Hagedorn phase~\cite{samir} and it would be interesting to study these possibilities in the future. With respect to inhomogeneities, near the bounce in the Hagedorn phase the sub-Hubble\footnote{By sub-Hubble we really mean here modes whose physical wave lengths are smaller than the time scale of the bounce.} thermal fluctuations remain small as the temperature which controls their amplitude is always less than the Planck scale, while the super-Hubble fluctuations freeze out before and are even smaller in amplitude. Thus inhomogenieties should also remain under control around the bounce in congruence with the BKL conjecture~\cite{bkl} according to which it is really the anisotropy rather than the inhomogeneity that one needs to worry about near the bounce. In any case, it is clear that the issue of anisotropy and inhomogenieties, both ``initial'' and their growth, demands a much more detailed analysis in the future.

From a technical point of view, we seem to require fine-tuning of the parameter $v_2$ which resulted from the fact that we did not want the modes to re-enter the Hubble radius before inflation could start. One may however be able to alleviate this problem significantly by considering  longer decay times, and just ensuring that in the Penultimate cycle enough radiation is produced so that it starts to dominate by the time the energy density increases to $\sim T_{\mth}^4$ in the contraction phase. In order to test these scenarios a more detailed numerical analysis will be required.

Finally, from the observational view point, it is clear that a very distinctive feature of the model is that the spectrum will make a transition from being nearly scale-invariant (when we see the modes that exited during contraction) to a very red spectrum at smaller length scales (which exit the Hubble radius during the power-law phase of inflation). This should manifest itself as a running in the spectral index, for which there is now some evidence~\cite{casas}. We do however need to make these predictions precise. Also, although it is clear that there are various effects which can give small corrections to scale-invariance of the modes that we observe in CMB\footnote{The biggest effect is expected due to deviations from a radiation dominated universe near the turnaround. If the turnaround temperature is close to the temperatures when the physically relevant modes that we observe today left the Hubble radius in the contracting phase, then we will see corrections to the scale-invariant spectrum. This seems a likely possibility for phenomenological reasons as discussed in section \ref{fine-tuning}.}, we  haven't yet investigated them in any detail to make connections to observations. We leave these issues for future.

\acknowledgements
 TB would like to thank Abhay Ashtekar,  Thorston Battefeld, Robert Brandenberger, Daniel Chung,  Anupam Mazumdar, John Peebles, Radu Roiban and Paul Steinhardt for several useful comments, criticisms  and suggestions. SA would like to thank Rob Caldwell and John Scwharz for useful discussions.
 \appendix
%%%%%%%%%%%%%%%%%%%%%%%%%%%%%%%%%%%%%%%%%%%%%%%%%%%%%%
\section{Phenomenological considerations for $\mu$}\label{muvalue}
It is known that in our present universe matter overtook radiation when the temperature was $\sim 100 ev$. After the phase transition from Hagedorn to radiation, which occurs $\sim T_H\sim 10^{-3}M_p$, the number of efoldings that had to evolve till matter-radiation equality is given by
\be
{a_{eq}\over a_p}={100 ev\over 10^{-3}M_p}\sim 10^{22}
\ee
Since the ratio $\rho_m/\rho_r$ also scales $\sim a$, it is clear that initially (at the transition) matter density has to be suppressed as compared to radiation by a factor
\be
\mu\sim 10^{-22}
\ee
The above estimate for $\mu$ is  only valid if the matter species under consideration is eventually going to be the dark matter. If it is some other unstable relic particle,  $\mu$ can obviously be much larger.
%%%%%%%%%%%%%%%%%%%%%%%%%%%%%%%%%%%%%%%%%%%%%%%%%%%%%%
\section{When can we ignore $\Om_c$ as compared to $\Om_r$?}\label{omc}
Since $\Om_r$ increases with entropy, to answer the above question let us first compute the smallest  $\Om_r$ possible in the emergent cyclic universe scenario. Clearly, this corresponds to the  $\Om_r$ for the critical cycle. Substituting $S\ra S_{cr}$ in (\ref{matching}), we have
\be
\Om_{r,\mt{min}}=\LF{4\over 3}\RF^4b_r^{4/3}\Om_k^2
\ee
Assuming $b_r\sim \cO(1)$ we therefore find that
\be
\Om_c\ll\Om_{r,\mt{min}}\im {\Om_c\over \Om_k^2}\ll 1
\ee

%%%%%%%%%%%

\end{document}